\documentclass[apjl]{emulateapj}
\usepackage{psfig,amsfonts,amsmath,graphicx,natbib,apjfonts}
\usepackage{ulem}

\begin{document}

\def\cfa{1}
\def\bonn{2}
\def\ta{3}
\def\uva{4}
\def\hart{5}
\def\york{6}

\title{Radio and X-ray Observations of the Type Ic SN\,2007gr Reveal an Ordinary, Non-relativistic Explosion}

\author{A.~M.~Soderberg\altaffilmark{\cfa}, A.~Brunthaler\altaffilmark{\bonn}, E.~Nakar\altaffilmark{\ta}, R.~A.~Chevalier\altaffilmark{\uva}, M.~F.~Bietenholz\altaffilmark{\hart,\york}}
    
\altaffiltext{\cfa}{Harvard-Smithsonian Center for Astrophysics, 60 Garden St., Cambridge, MA 02138, USA}
\altaffiltext{\bonn}{Max-Planck-Institute for Radio Astronomy, Auf dem Hugel 69, 53121 Bonn, Germany}
\altaffiltext{\ta}{Raymond and Beverly Sackler School of Physics \& Astronomy, Tel Aviv University, Tel Aviv 69978, Israel}
\altaffiltext{\uva}{University of Virginia, Astronomy Department, Charlottesville, VA 22904, USA}
\altaffiltext{\hart}{Hartebeesthoek Radio Observatory, P.O. Box 443, Krugersdorp, 1740, South Africa}
\altaffiltext{\york}{Dept. of Physics and Astronomy, York University, Toronto, M3J 1P3, Ontario, Canada}

\begin{abstract}
We present extensive radio and X-ray observations of the
nearby Type Ic SN\,2007gr in NGC\,1058 obtained with the Very Large
Array and the Chandra X-ray Observatory and spanning 5 to 150 days
after explosion.  Through our detailed modeling of these data, we
estimate the properties of the blastwave and the circumstellar
environment.  We find evidence for a freely-expanding and
non-relativistic explosion with an average blastwave velocity,
$\overline{v}\approx 0.2c$, and a total internal energy for the radio
emitting material of $E\approx 2\times 10^{46}$ erg assuming
equipartition of energy between electrons and magnetic fields
($\epsilon_e=\epsilon_B=0.1$).  The temporal and spectral evolution of
the radio emission points to a stellar wind-blown environment shaped
by a steady progenitor mass loss rate of $\dot{M}\approx 6\times
10^{-7}~\rm M_{\odot}~yr^{-1}$ (wind velocity, $v_w=10^3~\rm
km~s^{-1}$). These parameters are fully consistent with those inferred
for other SNe Ibc and are in line with the expectations for an
ordinary, homologous SN explosion.  Our results are at odds with those
of \citet{par10} who recently reported evidence for a relativistic
blastwave in SN\,2007gr based on their claim that the radio emission
was resolved away in a low signal-to-noise Very Long Baseline
Interferometry (VLBI) observation.  Here we show that the exotic
physical scenarios required to explain the claimed relativistic
velocity -- extreme departures from equipartition and/or a highly
collimated outflow -- are excluded by our detailed Very Large Array
radio observations.  Moreover, we present an independent analysis of
the VLBI data and propose that a modest loss of phase coherence
provides a more natural explanation for the apparent flux density loss
which is evident on both short and long baselines.  We conclude that
SN\,2007gr is an ordinary Type Ibc supernova.
\end{abstract}


\section{Introduction}
\label{sec:intro}

Nearly 25 years have elapsed since Type Ibc supernovae (SNe Ibc) were
first recognized as a distinct class of cosmic explosions
\citep{emn+85,fs85,wl85}.  They are now commonly understood to mark
the gravitational core-collapse of massive stars from which an
explosion launches ejecta to bulk velocities of $\sim
10,000~\rm km~s^{-1}$ (e.g.,~\citealt{fil97}).  Motivated by the
observed lack of hydrogen (and often helium) in their optical spectra,
the favored progenitors of SNe Ibc are Wolf-Rayet stars that have
ejected their massive envelope through strong stellar winds
\citep{bs86} or interaction with a binary companion \citep{pjh92}.
Recently, the observational realization that most long-duration
gamma-ray bursts (LGRBs) are accompanied by SNe Ibc has fueled a new
surge of interest in these massive star explosions (see \citealt{wb06}
and references therein).

The LGRB-SN connection and the relative rates of these events implies
that LGRBs are a rare subclass of SNe Ibc, distinguished by the
production of a relativistic outflow that decouples from the bulk SN
ejecta during explosion.  For many LGRBs, the energy associated with
the relativistic blastwave, $E\approx 10^{48}-10^{52}$ erg, is
comparable to that of the non-relativistic SN ejecta, $E_{\rm
SN}\approx 10^{51}-10^{52}$ erg (e.g.,~GRB\,030329/SN2003dh;
\citealt{bkp+03,hsm+03,mdt+03,fsk+05}).  This duality challenges the standard SN mechanism in which a homologous explosion couples at most $0.01\%$ of
the total energy to mildly-relativistic ejecta
\citep{tmm01}.  It is therefore generally accepted that a ``central
engine" --- an accreting and rapidly-rotating compact object
\citep{mwh01} or a magnetar (e.g.,~\citealt{tcq04}) --- is additionally required to power the energetic and relativistic blastwaves of LGRBs.  The
critical question is whether ordinary SNe Ibc also harbor central
engines, and in turn, the ability to produce even weak relativistic
outflows.

Identifying engine-driven, relativistic explosions requires
direct measurements of the blastwave velocity and energy.  While
optical SN emission predominantly traces the Nickel-56 synthesized in the bulk
SN ejecta \citep{arn82}, radio and X-ray observations directly probe the
synchrotron radiation produced as the blastwave shocks material in the
circumstellar medium (CSM;~\citealt{c82}).  Over the past decade,
dedicated radio studies of SNe Ibc have consistently pointed to
blastwave velocities of just $\langle v\rangle\approx 0.15c$ and
associated energies, $\langle E\rangle\approx 10^{47}$ erg, in agreement
with the expectations for an ordinary core-collapse explosion
\citep{c98,bkc02,bkf+03,skb+05,sck+06,cf06, ams07,sbp+08}.  Recently,
luminous radio emission was detected from the broad-lined Type Ic
SN\,2009bb pointing to an energetic ($E\gtrsim 10^{49}$ erg) and
relativistic ($v\gtrsim 0.85c$) outflow powered by a central engine
\citep{scp+10}.  This discovery marked the first relativistic SN
explosion identified without a detected gamma-ray counterpart.  Together
with the results of on-going radio supernova studies, this result
independently confirmed that the fraction of SNe Ibc with engine-driven
relativistic outflows is exceedingly small, $\sim 1\%$
\citep{bkf+03,snb+06,scp+10}.

Against this backdrop, \citet{par10} recently claimed evidence for a
mildly-relativistic blastwave ($v > 0.6c$) from the otherwise ordinary
Type Ic SN\,2007gr based on a low signal-to-noise Very Long Baseline
Interferometry (VLBI) radio observation.  Here we present an
independent analysis of these VLBI data and report in detail on our
extensive multi-frequency radio and X-ray observations of SN\,2007gr
obtained with the Very Large Array\footnote{The Very Large Array and
Very Long Baseline Array are operated by the National Radio Astronomy
Observatory, a facility of the National Science Foundation operated
under cooperative agreement by Associated Universities, Inc.}  (VLA)
and the Chandra X-ray Observatory (CXO).

In contrast to the hypothesis of \citet{par10}, our multi-frequency radio
data ($\nu_{\rm obs}=1.4-22$ GHz) and early X-ray observation reveal
that the SN emission is best explained by an ordinary,
non-relativistic blastwave with an average velocity,
$\overline{v}\approx 0.2c$, and an associated energy, $E\approx
2\times 10^{46}$ erg.  This result stems from the fact that we
directly measure the properties of the radio spectral peak across
multiple epochs.  We further demonstrate that a conservative
interpretation of the VLBI data is fully consistent with
non-relativistic SN expansion.  Moreover, we argue that the exotic
physical conditions required by a mildly relativistic outflow are
ruled out by our VLA observations of the temporal and spectral
evolution of the radio emission.  Overall, we find that the SN\,2007gr
blastwave properties are analogous to those of other Type Ibc
supernovae and starkly dissimilar from the relativistic jets of LGRBs.

\section{Observations}
\label{sec:obs}

SN\,2007gr was discovered by the Katzmann Automatic Imaging Telescope
on 2007 Aug 15.51 UT \citep{ml07} in NGC\,1058 ($d\approx 9.3$ Mpc;
\citealt{shm+96}).  An early spectrum of the supernova indicated a
Type Ibc classification based on the preliminary
identification of He I features \citep{cfl+07} although it has been
argued that the lines are better matched by C II motivating the
re-classification as a Type Ic SN \citep{vet+08}.  Based on the
non-detection of the SN in pre-discovery images from Aug 10.44 UT
\cite{ml07}, we adopt an explosion date of 2007 Aug $13 \pm 2$ days.

\subsection{Very Large Array}
\label{sec:vla}

We observed SN\,2007gr with the VLA beginning on Aug 17.41 UT ($\Delta
t\approx 4$ days after explosion) as part of the Very Large Array
Intensive Study of Naked Supernovae project (VISioNS;
\citealt{ams07}). We detected a radio source coincident with the
optical position with a flux density of $F_{\nu}=610\pm 40~\mu$Jy at
$\nu=8.46$ GHz \citep{sod07a}.  We subsequently initiated an intense VLA
follow-up campaign to study the temporal and spectral evolution of the
radio emission.

Radio data were collected at 1.43, 4.86, 8.46, and 22.5 GHz between
2007 August and 2008 January (Table~\ref{tab:vla}).  The bulk of our
data were obtained in the highest resolution VLA array (A array), and
all observations were taken in the standard continuum observing mode
with a bandwidth of $2\times 50$ MHz.  At 22.5 GHz we included
reference pointing scans to correct for the systematic 10-20 asec
pointing errors of the VLA antennas.  We used 3C48 (J0137+331) for
flux calibration, while phase referencing was performed using
calibrator J0254+395.  Data were reduced using standard packages
within the Astronomical Image Processing System (AIPS).  We fit a
Gaussian model to the radio SN in each observation to measure the
integrated flux density, and for our 8.46 GHz data set (our highest
resolution observations) we further report the associated centroid
position (Table~\ref{tab:vla}).

The statistical positional errors we infer from each epoch reflect
only the beam size and the signal-to-noise ratio of the SN detection
and they are dwarfed by the systematic errors introduced by the
atmosphere.  We determine a best estimate for the radio SN position
and associated uncertainty by calculating the weighted mean of the
centroid positions and find $\alpha\rm (J2000)=02^{\rm h}43^{\rm
m}27.9709{\rm s}, \delta\rm (J2000)=+37^{\rm o}20'44.692''$ with
uncertainties of 0.014 and 0.010 asec, respectively ($1\sigma$,
standard deviation of the mean).

\begin{figure}
\plotone{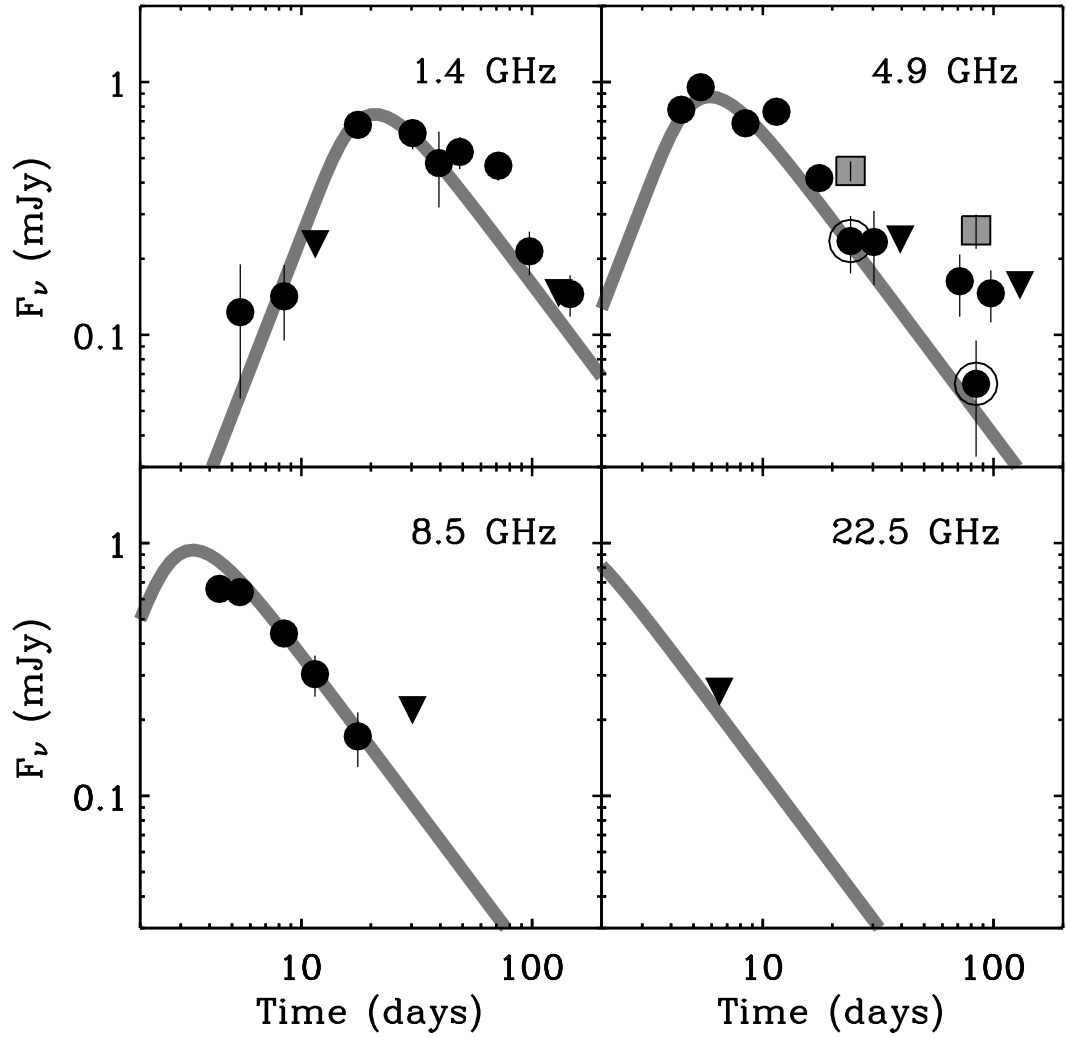}
\caption{Our radio light-curves of SN\,2007gr as obtained with the VLA at 
$\nu_{\rm obs}=1.4,~4.9,~8.5,$ and 22.5 GHz are well described by a
freely-expanding blastwave model in which a non-relativistic shock
accelerates electrons in a wind-stratified circumstellar environment
(grey curves; \S\ref{sec:dyn}).  At 4.9 GHz, the VLBI flux densities
at $\Delta t\approx 24$ and 84 days are shown as encircled black dots
(see \S\ref{sec:evn}).  The WSRT flux density measurements at $\Delta
t\approx 24$ and 84 days (grey squares) are discrepant with our nearly
simultaneous VLA measurements and could be due to
contamination by underlying host galaxy emission.}
\label{fig:lt_curve}
\end{figure}

The radio light curves are shown in Figure~\ref{fig:lt_curve},
spanning $\Delta t\approx 5$ to $150$ days.  The observations reveal an early
peak time of $\lesssim 5$ days at 8.46 GHz with an associated peak
spectral luminosity of $L_{\nu}\approx~9.7\times 10^{25}~\rm
erg~s^{-1}~Hz^{-1}$.  In comparison with the radio properties of other
SNe Ibc, SN\,2007gr is among the least luminous, about $10^3$ times
lower than SN\,1998bw (associated with GRB\,980425; \citealt{kfw+98}) and
SN\,2009bb \citep{scp+10} and most similar to SN\,2002ap \citep{bkc02}.

\subsection{Chandra X-ray Observations} 
\label{sec:cxo}

We additionally observed SN\,2007gr with the CXO
ACIS-I beginning on 2007 Aug 29.1 UT ($\Delta t\approx 16$ days) for
20 ksec \citep{sod07b}.  We do not detect an X-ray source coincident
with the optical and radio SN positions.  Adopting a power-law
spectral model with photon index, $\Gamma=2$, and a Galactic
foreground column density of $n_H\approx 3.7\times 10^{20}~\rm
cm^{-2}$, we place a $3\sigma$ upper limit on the X-ray flux of
$F_X\lesssim 3.9\times 10^{-15}~\rm erg~cm^{-2}~s^{-1}$ (0.2-10 keV).
At the distance of the SN, this implies a luminosity
of $L_X\lesssim 4.0\times 10^{37}~\rm erg~s^{-1}$ which is a factor of
$10^3$ below the afterglow luminosities of sub-energetic GRBs (e.g.,
GRB\,980425; \citealt{paa+00}) and similar to the X-ray luminosity of
SN\,2002ap on a comparable timescale \citep{scb+03}.
A comparison with our nearly simultaneous VLA observations indicates a
radio-to-X-ray spectral index steeper than $\approx {-0.7}$,
consistent with the measured values for other well-studied radio/X-ray
SNe Ibc shortly after explosion and during the epoch that inverse Compton
emission dominates the X-ray flux \citep{cf06}.

\section{A Simple Model for the Radio Emission}
\label{sec:model}

\begin{figure}
\plotone{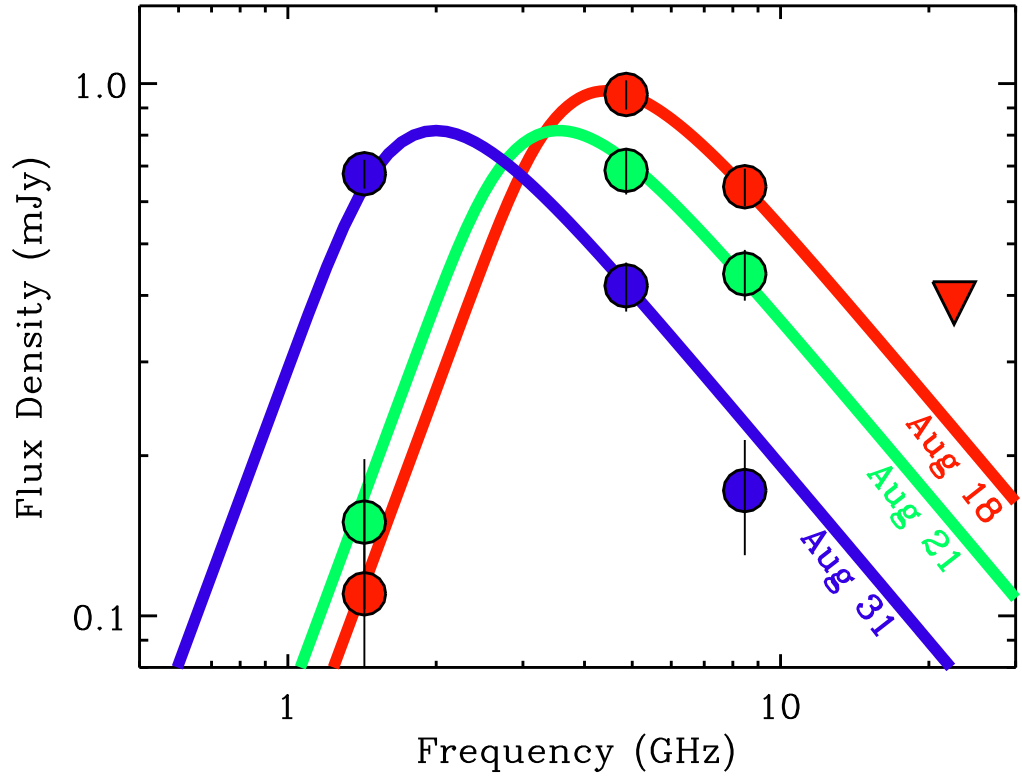}
\caption{The radio spectrum of SN\,2007gr across multiple epochs --  
$\Delta t\approx 5.4$ (red), 8.4 (green), and 17.6 (blue) days -- is well
described by a synchrotron self-absorbed spectral model with $F\propto \nu^{5/2}$ ($F_{\nu}\propto \nu^{-(p-1)/2}$) above (below) the spectral peak, $\nu_p$.
The observations indicate an electron energy index of $p\approx 3.2$.}
\label{fig:spectrum2}
\end{figure}

The radio emission from SNe, including those associated with LGRBs,
arises from the dynamical interaction of the blastwave with the
circumstellar medium (CSM; \citealt{c82,spn98}).  In this process, the
blastwave accelerates CSM electrons into a power-law distribution,
$N(\gamma)\propto \gamma^{-p}$, above a minimum Lorentz factor,
$\gamma_m$.  The interaction of the accelerated electrons with
amplified magnetic fields gives rise to non-thermal synchrotron
emission.  For SNe minimally affected by external absorption
processes, a low frequency spectral turn-over is produced by
synchrotron self-absorption (SSA) and defines the spectral peak
frequency, $\nu_p$.  In this scenario, the radio spectrum is given by
$F_{\nu}\propto \nu^{5/2}$ below $\nu_p$ and $F_{\nu}\propto
\nu^{-(p-1)/2}$ above $\nu_p$.  As shown in Figure~\ref{fig:spectrum2}, our
multi-frequency radio observations of SN\,2007gr are well-described by
a synchrotron self-absorbed spectrum with $p\approx 3.2$ across
multiple epochs.  We note that an extrapolation of the optically-thin
synchrotron spectrum to the X-ray band is consistent with the observed
CXO upper limit.

\subsection{Initial Constraints}
\label{sec:prelim}

For radio SNe characterized by SSA, the radius of the blastwave, $R$,
and its time-averaged velocity, $\overline{v}$, can be 
derived from the observed value of $\nu_p$ and the peak spectral
luminosity, $L_{\nu,p}$ \citep{c98,kfw+98}.  For $p\approx 3$, the
blastwave radius is given by 

\begin{equation}
R\approx 2.9\times 10^{16}\left(\frac{\epsilon_e}{\epsilon_B}\right)^{-1/19} \left(\frac{L_{\nu,p}}{10^{28}~\rm erg~s^{-1}~Hz^{-1}}\right)^{9/19}\left(\frac{\nu_p}{5~{\rm GHz}}\right)^{-1}~\rm cm
\label{eqn:radius}
\end{equation}

\noindent
where $\epsilon_e$ and $\epsilon_B$ are the relative fractions of
post-shock energy density shared by accelerated electrons and
amplified magnetic fields, respectively \citep{cf06}.  Throughout the paper
we make the assumption that the radio emitting
region is half of the total volume enclosed by a spherical
blastwave.  For SN\,2007gr, we find
$\nu_p\approx 4.0,~3.2,$ and 1.8 GHz and $L_{\nu,p}\approx
(9.8,~8.3,~8.3)\times 10^{25}~\rm erg~s^{-1}~Hz^{-1}$ at $\Delta
t\approx 5,~8,$ and 18 days (Figure~\ref{fig:spectrum2}), and thus
$R\approx (4.0,~4.7,~8.3)\times 10^{15}$ cm, respectively for
$\epsilon_e=\epsilon_B$.

The minimum internal energy, $E_{\rm min}$, required to power the
observed radio signal can be derived from the post-shock magnetic
energy density, $B^2/8\pi$.  As shown by
\citet{cf06}, the amplified magnetic field is directly determined from
the observed radio properties,

\begin{equation}
B\approx 0.43~\left(\frac{\epsilon_e}{\epsilon_B}\right)^{-4/19} \left(\frac{L_{\nu,p}}{10^{28}~{\rm erg~s^{-1}~Hz^{-1}}}\right)^{-2/19} \left(\frac{\nu_p}{5~{\rm GHz}}\right)~\rm G.  
\label{eqn:Bfield}
\end{equation}

\noindent
At $\Delta t\approx 5,~8$, 18 days the SN\,2007gr radio spectra imply
$B\approx 0.55,~0.45$, and 0.25 G. The minimum energy is

\begin{equation}
E_{\rm min}\approx 1.7\times 10^{44}~\left(\frac{B}{1~{\rm G}}\right)^2 \left(\frac{R}{10^{15}~{\rm cm}}\right)^3~\rm erg
\end{equation}

\noindent
where, by definition of $E_{\rm min}$ we have assumed
$\epsilon_e=\epsilon_B=0.5$. For our derived values of $R$ and $B$
across the three epochs, we infer $E_{\rm min}\approx
(3.2,~3.5,~6.0)\times 10^{45}$ erg.  A more realistic scenario allows
for a significant population of shocked protons such that
$(\epsilon_e=\epsilon_B)<0.5$.  The total internal energy is then
$E=(0.5/\epsilon_B)~E_{\rm min}$ and we find $E\approx
(1.6,~1.8,~3.0)\times 10^{46}$ erg for $\epsilon_B=0.1$.

In stark contrast, applying this same analysis to the observed radio
emission from nearby GRB-SNe 1998bw \citep{kfw+98,lc99} and 2006aj
\citep{skn+06} as well as SN\,2009bb
\citep{scp+10}, points unambiguously to mildly-relativistic outflows with $\overline{v}\sim c$ and with energies of $E\gtrsim 10^{49}$ erg.
The properties of the SN\,2007gr blastwave are clearly dissimilar from
those of relativistic explosions and are typical for ordinary,
non-relativistic SNe Ibc.


\subsection{A Dynamical Model}
\label{sec:dyn}

The temporal and spectral evolution of the radio emission can further
constrain the properties of the SN\,2007gr blastwave and local
environment. As the blastwave expands, the optical depth to SSA
decreases and, in turn, $\nu_p$ cascades to lower frequencies
producing the characteristic ``bell-shaped'' light-curves of radio SNe
Ibc \citep{c98}.  In modeling this evolution, we adopt the formalism of
\citet{skb+05} which assumes that the blastwave radius and the magnetic 
field evolve in time as power laws $R\propto t^{\alpha_r}$ and $B\propto
t^{\alpha_B}$ while $\epsilon_e$ and $\epsilon_B$ are constant
fractions of the post-shock energy density \citep{c98}.

We perform a global fit of the multi-frequency light-curves for four
parameters: $C_F$, $C_{\tau}$, $\alpha_r$ and $\alpha_B$ where $C_F$
and $C_{\tau}$ are normalization constants of the peak flux density
and optical depth at a reference epoch, $t_0$ (see
\citealt{skb+05} for a discussion of $C_F$ and $C_{\tau}$).  We find a
reasonable fit for $C_F\approx 3.7\times 10^{-50}~\rm g~s^{0.5}$,
$C_{\tau}\approx 3.0\times 10^{33}~\rm s^{3.6}$, $\alpha_r\approx 0.9$
and $\alpha_B\approx -1$ at $t_0=10$ days.  The resulting model fits
are shown in Figure~\ref{fig:lt_curve} and imply $R\approx 5\times
10^{15}~(t/10~\rm days)^{0.9}~\rm cm$ (i.e.,~$\overline{v}\approx
0.2c~(t/10~\rm days)^{-0.1}$) and $B\approx 0.4~(t/10~\rm
days)^{-1}~\rm G$.  Thus, the radio observations point to a
freely-expanding SN in which the bulk ejecta travel with a constant
velocity while the post-shock material is slightly decelerated.  With
these constraints, the the total internal energy of the radio emitting
material is $E\approx 2\times 10^{46}~(t/10~\rm days)^{0.7}~\rm erg$
for $\epsilon_e=\epsilon_B=0.1$. These scalings are fully consistent
with our preliminary estimates from individual-epoch analysis of the
radio spectra in \S\ref{sec:prelim}.  

\begin{figure}
\plotone{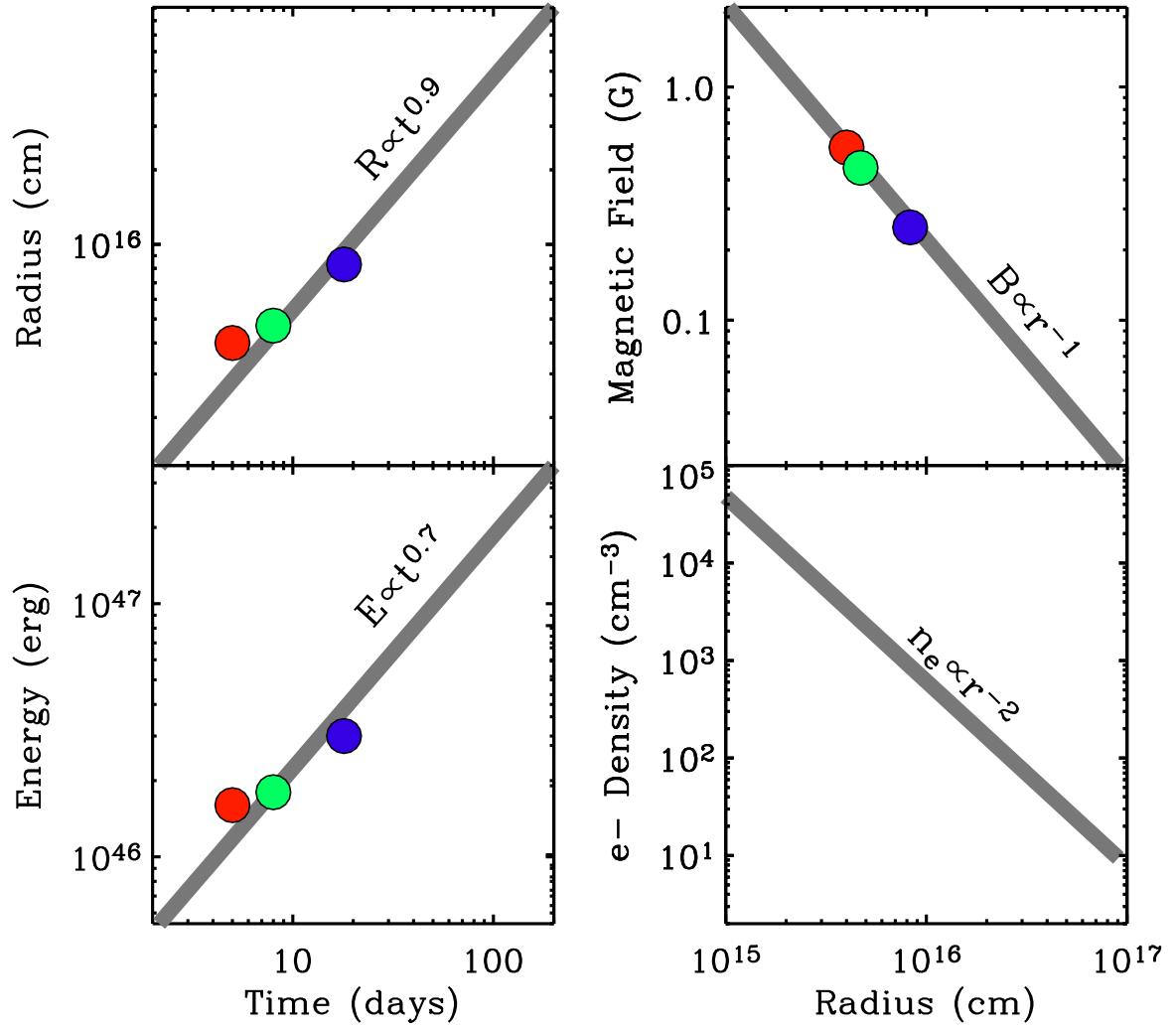}
\caption{The temporal and radial evolution of the physical parameters 
associated with the SN\,2007gr radio emitting material are shown in
grey as derived from our dynamical model fit to the VLA data (see
Figure~\ref{fig:lt_curve}).  For comparison, we show the individual
data points for the blastwave radius, energy, and magnetic field
intensity as inferred from a single-epoch analysis of the SSA radio
spectra at $\Delta t\approx 5$ (red), 8 (green) and 18 (blue) days
(Figure~\ref{fig:spectrum2}). These individual points are overall
consistent with our freely-expanding and non-relativistic dynamical
model characterized by a wind-stratified circumstellar medium. }
\label{fig:params}
\end{figure}

Next, the electron number density is
$n_e=(p-2)/(p-1)~(B^2/8\pi\gamma_m m_e c^2)\approx 2\times
10^3~(r/5.5\times 10^{15}~\rm cm)^{-2}~\rm cm^{-3}$ with
$\gamma_m\approx 2$ (here we maintain the assumption that $p\approx 3$
and $\epsilon_e=\epsilon_B=0.1$).  The density profile of the
circumstellar environment is thus consistent with the expectations for
a stellar wind with a constant mass loss rate and wind speed.
Assuming a typical Wolf-Rayet wind velocity of $v_w=10^3~\rm
km~s^{-1}$ (e.g.,~\citealt{cgv04}), and a nucleon-to-electron ratio of
two (appropriate for a predominantly helium stellar wind), we infer a
mass loss rate of $\dot{M}=4\pi n_e m_p R^2 v_w \approx 6\times
10^{-7}~\rm M_{\odot}~yr^{-1}$ for the SN\,2007gr progenitor star.
The temporal evolution of the physical parameters associated with the
SN\,2007gr blastwave and the circumstellar material are displayed in
Figure~\ref{fig:params}.  We emphasize that the blastwave dynamics and
physical properties that we derive for SN\,2007gr are similar to those
of ordinary radio SNe Ibc (\citealt{bkf+03,ams07} and references
therein) and fully consistent with dynamical models for radio SNe Ibc
\citep{c98,cf06}.

Finally we note that an independent constraint on the partition
fractions can be obtained from the blastwave velocity since the
observed synchrotron radiation requires a sufficient energy to
accelerate electrons to relativistic speeds.  For a non- or
mildly-relativistic shock, the requirement that $\gamma_m > 1$ implies

\begin{equation}
\gamma_m\approx 460~\epsilon_e \left(\frac{v}{c}\right)^2 \left(\frac{p-2}{p-1}\right)> 1
\label{eqn:gamma_m}
\end{equation}

\noindent
which results in a lower limit of $\epsilon_e
\gtrsim 0.44~(v/0.1c)^{-2}$ for $p\approx 3$ (\citealt{cf06}; see also
\citealt{skb+05}). For SN\,2007gr, this constraint leads to 
$\epsilon_e \gtrsim 0.1$ for $\overline{v} \gtrsim 0.2c$ and is fully
consistent with our equipartition model (\S\ref{sec:dyn}), therefore
supporting our derived energy estimate.

\subsection{Bulk Ejecta Parameters}

Based on modeling of the optical light-curves and spectra,
\citet{hvk+09} reported values for the total kinetic
energy and mass of the SN ejecta of $E_{\rm SN}\approx (1-4)\times
10^{51}~\rm erg$ and $M_{\rm ej}\approx (2-3.5) M_{\odot}$,
respectively.  The velocity of the bulk ejecta is thus $v_{\rm
bulk}\approx (1.8~E_{\rm SN}/M_{\rm ej})^{1/2}\approx
9,000-13,600~\rm km~s^{-1}$ \citep{inm+03}, in line with the observed
photospheric velocities \citep{vet+08,hvk+09}.  Theoretical
considerations predict that the coupling of energy and
velocity within the homologous-expanding SN ejecta is
characterized by 

\begin{equation}
E(v)\approx 3.7\times 10^{47} \left(\frac{E_{\rm SN}}{10^{51}}\right)^{3.59} \left(\frac{M_{\rm ej}}{M_{\odot}}\right)^{-2.59} \left(\frac{v}{0.1c}\right)^{-5.18}~\rm erg
\label{eqn:energy}
\end{equation}

\noindent
\citep{mm99, bkc02}.  For SN\,2007gr, the bulk ejecta parameters predict
that the ejecta traveling at $v\ge 0.2c$ carry an energy of $E\approx
(0.1-4.5)\times 10^{46}~\rm erg$, which is fully consistent with the
energy inferred from our free-expansion model (\S\ref{sec:dyn}).
Therefore there is sufficient energy in the high-velocity ejecta to
account for the observed radio signal within the framework of a
standard homologous explosion.

\section{A Comparison with {\it Paragi et al. 2010}}
\label{sec:comp}

As detailed in the previous section, the temporal and spectral
evolution of the SN\,2007gr radio emission point to an ordinary and
non-relativistic blastwave with velocity, $\overline{v}\approx 0.2c$,
and energy, $E\approx 2\times 10^{46}$ erg, for
$\epsilon_e=\epsilon_B=0.1$.  Similar blastwave parameters are
independently inferred from the bulk ejecta properties and expected
energy profile of the ejecta (Eqn.~\ref{eqn:energy}).  In stark
contrast, \citet{par10} propose a relativistic blastwave velocity of
$\overline{v}\gtrsim 0.6c$ based on their analysis of a low
signal-to-noise VLBI observation of the SN at $\Delta t\approx 84$
days from which they claim a lower limit on the blastwave radius.
 of $R\gtrsim 1.3\times 10^{17}$ cm.  This is a factor of $\gtrsim 3$ larger
than our modeling estimate.  In the following sections we address
this claim by first considering the implications of a
mildly-relativistic outflow in the framework of our dynamical model
for the SN\,2007gr radio light-curves and spectra.  Following this
discussion, we present the results from our independent analysis of
the VLBI data.

\subsection{Severe Departures from Equipartition?}
\label{sec:equip}

\begin{figure}
\plotone{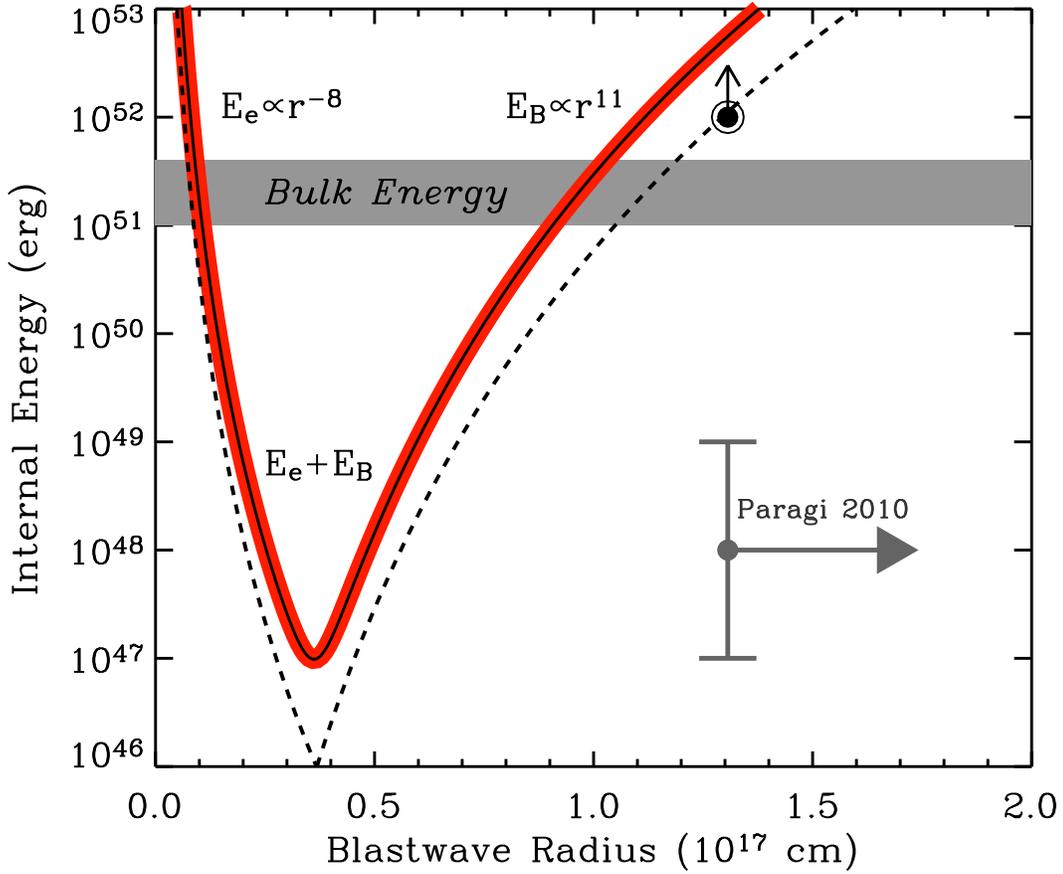}
\caption{The total internal energy of the SN\,2007gr radio emitting material 
at $\Delta t\approx 84$ days after explosion. Our modeling of the SSA
radio spectrum across multiple epochs provides a tight constraint on
the time-averaged velocity of the ejecta, $\overline{v}\approx 0.2c$,
and a minimum internal energy of $E_{\rm min}\approx 6\times 10^{46}$
erg at $\Delta t\approx 84$ days (\S\ref{sec:dyn}) for
$\epsilon_e=\epsilon_B=0.5$ (dashed black lines). Assuming these
components each account for just 10 percent of the total internal
energy, we find $E\approx 2\times
10^{47}$ erg for $\epsilon_e=\epsilon_B=0.1$ (solid red line).  Deviations
from equipartition imply faster velocities for blastwaves
energetically dominated by magnetic fields.  Such deviations are also
associated with a steep increase in the total energy budget.  The
mildly-relativistic blastwave properties reported by \citet{par10}
(grey arrow) require extreme departures from equipartition, and assuming
a spherical outflow implies a {\it minimum}
internal energy of $E\approx 10^{52}$ erg for the blastwave (black encircled
dot and arrow).  In this extreme scenario, the blastwave energy would rival the
bulk explosion energy of $E_{\rm SN}\approx (1-4)\times 10^{51}$ erg
(horizontal grey bar) as estimated from modeling of the optical data
\citep{hvk+09}.   As discussed in \S\ref{sec:evn}, the VLBI observation
suffers from a low signal-to-noise ratio and likely systematic effects
which casts doubt on the claimed mildly relativistic expansion velocity.} 
\label{fig:energy_radius_07gr}
\end{figure}

As shown in Equation~\ref{eqn:radius}, the blastwave radius depends
only weakly on the partition fractions, $R\propto
(\epsilon_e/\epsilon_B)^{-1/19}$.  This stems from the fact that the
energy in electrons ($E_e$) and amplified magnetic fields ($E_B$)
scale with the spherical blastwave radius as $E_B\propto
\overline{R}^{11}$ and $E_e\propto \overline{R}^{~-8}$, respectively,
such that the total energy budget ($E_e+E_B$) is minimized at
equipartition (Figure~\ref{fig:energy_radius_07gr}).  Therefore, in
order to accommodate a factor of $\gtrsim 3$ increase in the
time-averaged velocity within the framework of our spherical blastwave
model would require severe departures from equipartition,
$\epsilon_e/\epsilon_B\lesssim 10^{-9}$.  Such deviations are
unprecedented in astrophysical systems and indeed, detailed VLBI
studies of other radio SNe that point to relative partition fractions
close to equipartition (e.g., $\epsilon_e/\epsilon_B\approx 0.004$ for
SN\,1993J;
\citealt{fb98,crb04}) while broadband modeling of GRB afterglows typically
indicates $\epsilon_e/\epsilon_B\approx 10$ \citep{pk02,yhs+03}.

Furthermore, Eqn.~\ref{eqn:gamma_m} shows that a $\overline{v}\gtrsim
0.6c$ outflow requires the relativistic electrons to harbor a
significant fraction of the post-shock energy density, $\epsilon_e
\gtrsim 0.01$. This is a factor of $10^7$ higher than the value
required to accommodate the VLBI measurement within our spherical
blastwave model.  To reconcile this inconsistency would require
an atypical modification of the electron energy distribution.

Severe deviations from equipartition also impose a significant
increase in the total energy of the radio emitting material.  In
\S\ref{sec:dyn} we report the modest energetics required by a
non-relativistic model for the blastwave in equipartition, $E\approx
2\times 10^{46}$ erg.  However, a mildly-relativistic velocity of
$\overline{v}\gtrsim 0.6c$ would require a magnetically dominated
blastwave with an enormous energy of $E\gtrsim 10^{52}$ erg, exceeding
the total bulk energy of the explosion
(Figure~\ref{fig:energy_radius_07gr}).

\subsection{A Collimated Outflow?}
\label{sec:jets}

To avoid severe departures from equipartition, \citet{par10} propose
that the outflow is highly collimated into jets with opening angles of
just $\theta_j\approx 15$ degrees.  Thus the radio emitting region
fills only a fraction of the total solid angle, $f_b\equiv(1-{\rm
cos}\theta_j)\approx 0.03$.  Reducing the area of the radio emitting
region serves to increase the radius associated with equipartition,
thus bringing the minimum of the ($E_e+E_B$) curve closer to their
claimed blastwave radius.  In this scenario, they adopt a blastwave
energy of $E\approx 3\times 10^{47}$ erg by allowing for modest
departures from equipartition.  The associated isotropic-equivalent
energy would be, $E_{\rm iso}\equiv (E/f_b)\approx 10^{49}$ erg.
\citet{par10} further adopt a mass loss rate of $\dot{M}\approx
3\times 10^{-7}~M_{\odot}$, a factor of two lower than our own
estimate.

Such relativistic and highly collimated jets are are incompatible with
the standard expectations for a homologous SN explosion. Thus,
\citet{par10} appeal to a model in which the blastwave
detached from the bulk ejecta at the time of explosion, similar
to the model for GRB-associated SNe.  In this scenario, the
trans-relativistic SN\,2007gr jets raced ahead of the bulk spherical
outflow and freely-expanded\footnote{Here we note that
trans-relativistic outflows may not follow formal Blandford-McKee
dynamics owing to their low bulk Lorentz factors.} until they swept up
a circumstellar mass comparable to their own rest mass, causing a
deceleration to non-relativistic expansion at a time, $t_{\rm dec}$
\citep{wkf98}.  Thereafter, the dynamics of the outflow approach the
Sedov-Taylor solution characterized by a radial expansion, $R\propto
t^{2/3}$, into a wind environment (e.g.,~\citealt{sed46}).  Due to
lateral spreading of the jets, the outflow also approaches spherical
symmetry within a dynamical timescale of $t_{\rm dec}$. This is roughly the
time for the outflow radius to double in size, and thus $t_{\rm
sph}\approx 2^{3/2}~t_{\rm dec}$.

Throughout its evolution leading up to $t_{\rm sph}$, the
blastwave experiences several dynamical phase transitions, from
freely-expanding to decelerated expansion, relativistic to
non-relativistic, and collimated to spherical evolution. These
transitions give rise to abrupt changes in the temporal decay of the
radio light-curves (e.g.,~\citealt{sph99,fwk00}). In particular,
spherical and non-relativistic expansion requires a steepening of the
optically thin flux density to $F_{\nu}\propto t^{-2.7}$ for $p\approx
3$ \citep{cl00}.  In the case of SN\,2007gr, however, the optically
thin flux evolution decays steadily throughout our radio observations
spanning $\Delta t\approx 5-150$ days: $F_{\nu}\propto t^{-0.9\pm
0.2}$ (1.4 GHz), $F_{\nu}\propto t^{-0.8\pm 0.2}$ (4.9 GHz), and
$F_{\nu}\propto t^{-1.0\pm 0.1}$ (8.5 GHz; see
Figure~\ref{fig:lt_curve}).  Thus we find no evidence for any phase
transitions on this timescale, and we highlight that the
observed temporal decay is statistically inconsistent with the
expectations for a decelerated blastwave.

The timescale required for the jets to reach $t_{\rm dec}$ depends on
the rate at which the jets spread sideways. We consider two extreme
cases that bracket the range of hydrodynamic evolutions: (i) lateral
spreading with a rate, $v\sim c$, and (ii) minimal spreading (see
\citealt{grl05} for a full discussion).  In the former case, 
$t_{\rm dec}\approx 100~(E/10^{51}~\rm erg)(\dot{M}/10^{-5}~\rm
M_{\odot}~yr^{-1})^{-1}$ days, while the latter case is longer by a
factor of $f_b^{-1}$ since it assumes there is no lateral jet
spreading during relativistic expansion \citep{cl00,wax04}.  For
\citet{par10}'s proposed jet parameters, a SN\,2007gr relativistic
outflow would decelerate at $t_{\rm dec}\approx 1-30$ days and
approach spherical symmetry at $t_{\rm sph}\approx 3-90$ days.
Therefore, even in the unlikely scenario that the
SN\,2007gr radio outflow was initially jetted, relativistic, and
experienced minimal lateral spreading, by the epoch of the VLBI
observation ($\Delta t\approx 84$ days) these ejecta would have
roughly transitioned to non-relativistic and spherical expansion.
Moreover, as noted above, the observed light-curves show no evidence
for such phase transitions on these timescales.  

In summary, the exotic physical scenarios required by a
mildly-relativistic blastwave with a velocity of $\overline{v}\gtrsim
0.6c$ proposed by \citet{par10} are overall inconsistent with the
observed temporal and spectral evolution of the SN\,2007gr radio
emission.  We also show that a homologous free-expansion blastwave
model with a non-relativistic velocity, $\overline{v}\approx 0.2c$ and
an energy, $E\approx 2\times 10^{46}$ erg, characterized by shock
micro-physics near equipartition provide an excellent (and more natural)
description of the radio observations.  We next turn to an independent
analysis of the SN\,2007gr VLBI dataset.

\subsection{Very Long Baseline Interferometry Data}
\label{sec:evn}

SN\,2007gr was observed with the European Very Long Baseline
Interferometry Network (EVN) on two epochs beginning at 2007 Sep 6 and
Nov 5 UT ($\Delta t\approx 24$ and 84 days, respectively).  We
retrieved the data from the EVN
archive\footnote{http://archive.jive.nl/scripts/portal.php} (programs
RP007 and GP044; PI Paragi).  Both EVN observations were carried out at 4.9
GHz.  The first observation was carried out in real-time e-VLBI mode,
lasted 12 hours, and included the Darnhall, Medicina, Jodrell Bank,
Onsala, Torun, and Westerbork (WSRT) antennas.  The data were recorded
with four 8 MHz sub-bands in dual polarization and 2-bit sampling,
resulting in an aggregate data rate of 256 Mbps. During the
phase-referencing cycles, one minute was spent on the reference source
J0253+3835, and 4.5 minutes on SN\,2007gr. After 2 cycles including
SN\,2007gr, two additional background quasars (J0230+4032 and
J0247+3254) were also observed while 3C\,454.3 and 3C\,84 were used as
fringe finders.

The second observation included the Medicina, Jodrell Bank, WSRT,
Cambridge, Torun, Noto, Onsala, Effelsberg, Hartebeestoek, and the Green
Bank Telescope antennas.  The observation lasted 10 hrs with a recording
rate of 1 Gbps for the EVN stations (eight sub-bands of 16 GHz in dual
polarization and 2-bit sampling), while the GBT observed with 512 Mbps
(same setup, but with 1-bit sampling). The observing scheme was
similar to the first epoch, but the two background quasars were
observed less frequently (only after 4 cycles on SN\,2007gr).

The original analysis of these data was reported by \citet{par10};
here we present an independent analysis. The data reduction was
performed using standard packages within AIPS. Total electron content
maps of the ionosphere were used to correct for associated phase changes.
A-priori amplitude calibration was applied using system
temperature measurements and standard gain curves. We performed
phase-calibration using the data for 3C\,84 to remove instrumental
phase offsets among the frequency bands. We then fringe-fitted the
data from J0253+3835 and transferred the phase calibration to
SN\,2007gr.

\subsubsection{VLBI Epoch 1}
\label{sec:e1}

\begin{figure}
\plotone{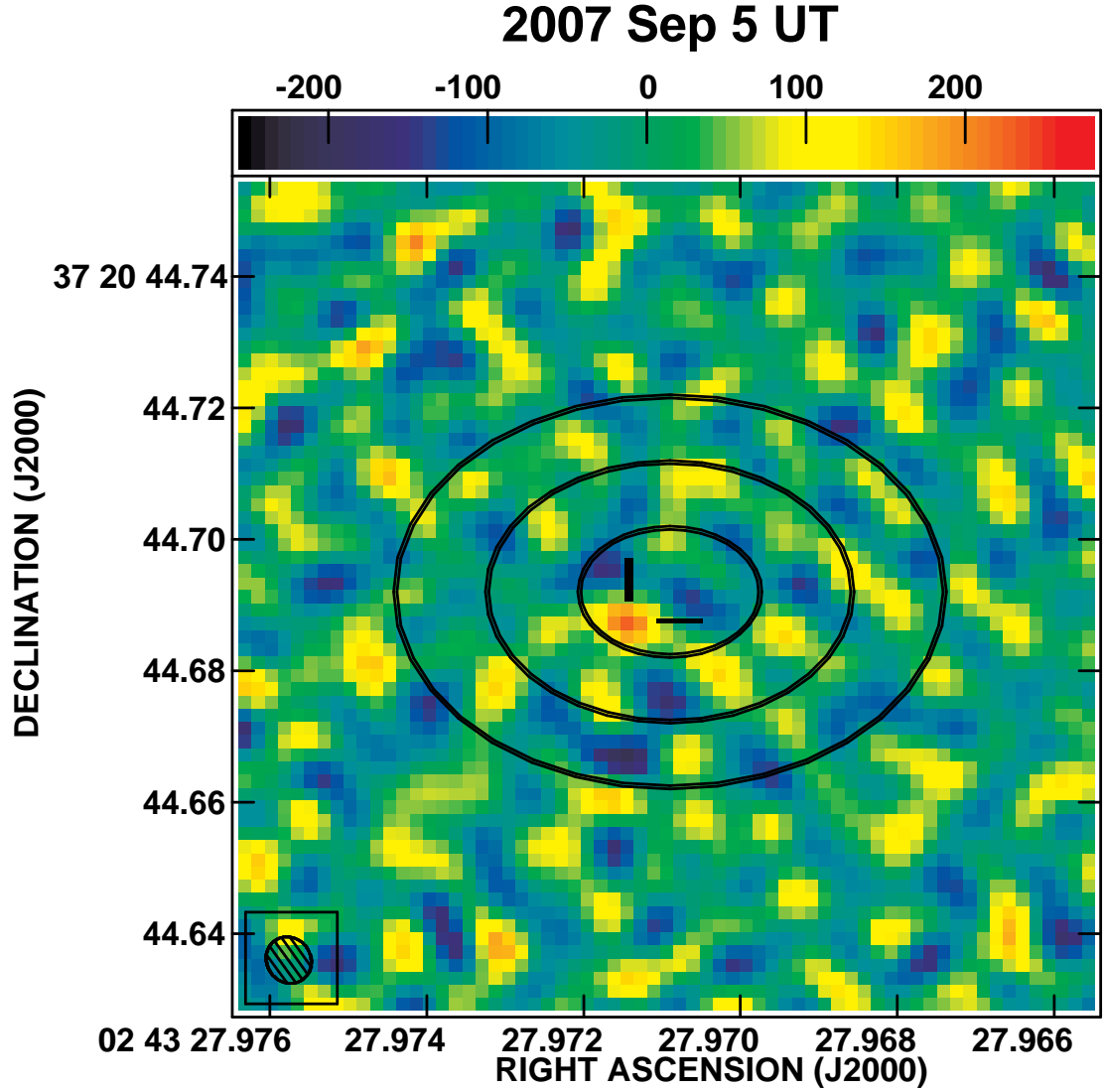} 
\caption{EVN observation of SN\,2007gr on 2007 Sep 6 UT.  The color scale
extends from $F_{\nu}=-250$ to $270~\mu$Jy per beam.  The
uncertainty associated with our weighted mean VLA position is shown as
concentric error ellipses corresponding to 1$\sigma$, 2$\sigma$, and
3$\sigma$.  Within the $1\sigma$ error ellipse, we detect an
unresolved source with flux density $F_{\nu}=235\pm60~\mu$Jy which we
identify as the SN (tick marks). }
\label{fig:EPOCH1}
\end{figure}

In the first VLBI epoch, we confirm the detection of an unresolved
source at coordinates, $\alpha\rm (J2000)=02^{\rm h}43^{\rm
m}27.9715^{\rm s}, \delta\rm (J2000)=+37^{\rm o}20'44.687''$ ($\pm 5$
mas in each coordinate, dominated by the positional uncertainty of the
phase calibrator) which is within $2\sigma$ of our weighted mean VLA
SN position .  We measure an integrated flux density of
$F_{\nu}=235\pm 60~\mu$Jy for the source and note that this is the
strongest source detected within the $3\sigma$ VLA localization region
(Figure~\ref{fig:EPOCH1}). The likelihood of detecting a source with
S/N$\approx 3.9$ (rms noise$=60~\mu$Jy) in one one of the 26
independent VLBI beams within the $3\sigma$ position ellipse is
roughly $3\times 10^{-3}$.  This, along with the consistent flux of
the source with our nearly coincident VLA measurement ($F_{\nu,
4.86~\rm GHz}=233\pm 76~\mu$Jy at Sep 12.31 UT; Table~\ref{tab:vla})
supports the identification of the VLBI source as SN\,2007gr.  We note
that \citet{par10} reported a higher integrated flux density for the
SN by a factor of $1.8\pm 0.6$ while their associated rms noise level
was also higher by a factor of 1.3.

Since the SN emission is unresolved, the beam size ($7.3 \times 6.7$
mas) implies an upper limit on the diameter of the emitting region of
$\lesssim 9.3\times 10^{17}~\rm cm$.  Assuming spherical expansion,
this constrains the time-averaged velocity to be $\overline{v}\lesssim
7.4c$.

\subsubsection{VLBI Epoch 2}
\label{sec:e2}

In the second epoch, the longest baselines provided by the Green Bank
Telescope and Hartebeesthoek antennas result in a higher resolution
image with a synthesized beam of $3.0 \times 0.93$ mas (natural
weighting).  In their original analysis of these data,
\citet{par10} reported the weak detection of a source coincident with
the first epoch SN position with flux density, $F_{\nu}\approx
60~\mu$Jy (rms map noise, $13~\mu$Jy).  In comparison, coincident
observations at $\nu_{\rm obs}=4.9$ GHz with the WSRT and VLA indicate
somewhat higher SN flux densities of $F_{\nu,4.9~\rm GHz}=259\pm
40~\mu$Jy \citep{par10} and $146\pm 34~\mu$Jy (Table~\ref{tab:vla}),
respectively.  \citet{par10} interpreted the WSRT-VLBI flux
discrepancy to the radio emission being resolved on the
{\it longest} baselines.  Adopting the high-resolution beam size as
a lower limit on the angular diameter, they claimed that the
apparently resolved SN emission (i.e., ``missing flux'') would require a
time-averaged blastwave velocity of $\overline{v}\gtrsim 0.6c$.  We
note, however, that the VLA-VLBI flux discrepancy is less significant
(consistent at the $2\sigma$ level) and we further address the VLA-WSRT
flux discrepancy at the end of this section in the context of host
galaxy contamination.

\begin{figure}
\plotone{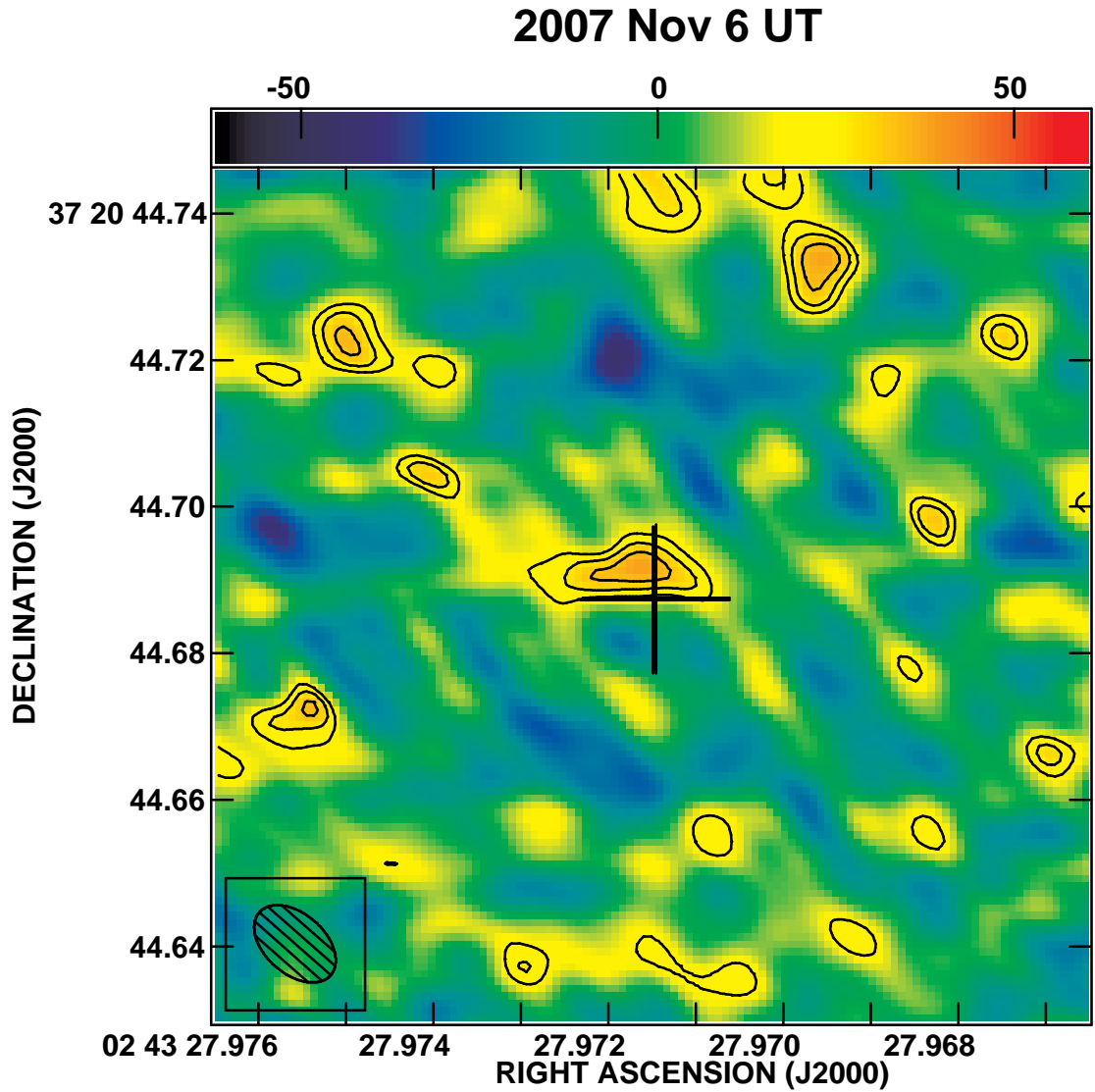} 
\caption{EVN observation of SN\,2007gr on 2007 Nov 5 UT, excluding the Green Bank and Hartebeesthoek antennas. The color scale
extends from $F_{\nu}=-60$ to $60~\mu$Jy per beam and the contours represent
1.5$\sigma$, 2$\sigma$, and 2.5$\sigma$ levels (rms noise is
$13\mu$Jy per beam).  At the position of the SN\,2007gr measured in the first
VLBI epoch (black cross; $2\sigma$ errors) we detect a weak and
source with integrated flux density, $F_{\nu}=64\pm
31~\mu$Jy.  Fitting a Gaussian in the image plane suggests that the emission
region is {\it apparently} marginally extended with a major axis of
$20^{+8}_{-10}$ mas, an upper limit of $\lesssim 9.9$ mas ($3\sigma$)
for the minor axis, and a position angle of $\theta=98^{+13}_{-10}$
degrees after convolution with the restoring beam.   We attribute this 
apparent extension to the low signal-to-noise ratio of the image.}
\label{fig:EPOCH2}
\end{figure}

In our independent analysis of the second epoch VLBI data, we
attempted to verify the source extension by excluding the data from
the Green Bank and Hartebeesthoek antennas resulting in a lower
resolution image with a synthesized beam size of $13\times 8.2$ mas,
comparable to that in the first VLBI epoch.  We confirm the presence
of a weak and {\it apparently} extended source consistent with the
position of the SN and measure an integrated flux density of
$F_{\nu}\approx 64~\mu$Jy (rms noise level, $\sigma=13~\mu$Jy per
beam, Figure~\ref{fig:EPOCH2}).  Thus, a flux discrepancy persists
even on shorter baselines. We note that \citet{par10} similarly reported
no increase in the flux density when the data from the Hartebeesthoek and
Greenbank antennas were excluded. If this effect is attributed to
resolved emission from a relativistic outflow, the required velocity
is significantly higher, $\overline{v}\gtrsim 2.6c$, which implies far
more stringent requirements on the properties and dynamics of the
blastwave than \citet{par10}'s proposed value of $\overline{v}\gtrsim
0.6c$ (see \S\ref{sec:comp}).

An alternative interpretation is that the observations suffer from
a modest loss of phase coherence, a common phenomenon in
phase-referenced VLBI observations which is difficult to correct at low
signal-to-noise ratios.  Such losses cause an apparent decrease in
the total flux density \citep{mrp+10},
and could explain the modest flux discrepancy observed
between the nearly coincident VLBI-VLA measurements\footnote{Indeed,
\citet{par10} similarly report that the data were ``affected by modest
phase errors''.}.  We therefore propose a more conservative
interpretation in which the flux discrepancy and apparent extension of
the SN -- revealed even on shorter VLBI baselines -- are the result of
modest phase decoherence affecting this low signal-to-noise observation.
Under this interpretation, we adopt the beam size as an {\it
upper limit} on the apparent diameter of the otherwise unresolved SN
emission to place a constraint on the blastwave velocity of
$\overline{v}\lesssim 2.6c$ which is fully consistent with our
non-relativistic spherical blastwave model.

Finally, we consider the modest flux discrepancy observed for the
VLA and WSRT measurements near the epoch of the second VLBI
observation.  Given the that the WSRT synthesized beam (3.2 asec) is
significantly larger than that of our VLA observation (1.4 asec,
B-array) we propose that the WSRT data may be contaminated by diffuse
host galaxy emission.  To test this hypothesis, we analyzed VLA
observations of the host galaxy, NGC\,1058, obtained at 4.9 GHz prior
to the discovery of SN\,2007gr and publicly available from the NRAO
archive\footnote{http://archive.cv.nrao.edu/}.  We find that the SN is
located in a strongly star-forming region of its host galaxy
characterized by enhanced radio emission.  From these data, we measure
diffuse host galaxy emission at the position of the SN with a flux
density per beam of $F_{\nu}\approx 185\pm 15~\mu$Jy (D-array, 15.4
asec beam) and $F_{\nu}\approx 23\pm 29~\mu$Jy (B-array, 1.4 asec
beam).  Given that the WSRT resolution is intermediate between these
two cases, it is plausible that the WSRT observations are contaminated
by low-level and diffuse host emission.  This is further supported by
the fact that an earlier WSRT observation at $\Delta t\approx 24$ days
showed a similarly high flux density in comparison to our nearly 
contemporaneous VLA observations (see Figure~\ref{fig:lt_curve}).

In conclusion, we attribute the modest VLA-VLBI flux discrepancy and
apparent extension of the SN emission in the low S/N second epoch VLBI
observation to the likely loss of phase decoherence.  We further
attribute the VLA-WSRT flux discrepancy to contamination by diffuse
host galaxy emission near the explosion site.

\section{Conclusions} 
\label{sec:conc} 

In this paper, we present a critical analysis of blastwave properties
of Type Ic SN\,2007gr following the recent claim by \citet{par10} that
this otherwise ordinary SN produced a relativistic outflow similar to
those of nearby LGRBs.  We show that the full dataset of radio (VLA
{\it and} VLBI) and X-ray observations for SN\,2007gr are more
naturally explained by an ordinary, non-relativistic, and homologous
SN explosion.  Our conclusions stem from the fact that we directly
measured the frequency and flux density of $\nu_p$ across several
epochs.  This result underscores the necessity of multi-frequency and
long-term radio monitoring of SNe Ibc in the search for relativistic
outflows.  We conclude with the following points:

\noindent
\begin{enumerate}
\item{A freely-expanding and non-relativistic blastwave model best reproduces
our extensive radio and X-ray observations, indicating an expansion
velocity of $\overline{v}\approx 0.2c$ and energy of
$E\approx 2\times 10^{46}$ erg.  These blastwave parameters are
consistent with those inferred from the bulk ejecta
properties given the expected energy profile of the
ejecta.}

\noindent
\item{\citet{par10}'s proposed mildly-relativistic blastwave velocity would
require exotic physical scenarios (severe departures from
equipartition or a decelerated, detached blastwave) that are implausible
given the requirements of our VLA observations.}

\smallskip

\noindent
\item{Through our independent analysis of the VLBI data sets, we confirm the 
weak detection of a source consistent with our VLA position for
SN\,2007gr.  However, in our conservative interpretation of the low
signal-to-noise second epoch detection, we attribute the apparent loss
in flux density to modest phase decoherence, instead of relativistic
SN expansion.  We also suggest that the WSRT observation is
contaminated by diffuse emission from the host galaxy.}

\end{enumerate}

Finally, we note that while the data for SN\,2007gr do not point to a
relativistic explosion, they do offer new insight on the non-thermal
properties of Type Ibc SNe.  In addition to being one of the {\it
nearest} of such explosions discovered to date, SN\,2007gr is also one
of the {\it least} radio luminous.  This can be directly attributed to
its low density circumstellar environment that was shaped by a steady
mass loss rate, $\dot{M}\approx 6\times 10^{-7}~\rm
M_{\odot}~yr^{-1}$ (\S\ref{sec:dyn}).  For comparison, this is
$10^3-10^4$ times lower than the mass loss rates inferred for radio
SNe Ibc with the strongest circumstellar interaction (e.g.,~SNe 2003L,
2003bg;~\citealt{skb+05,sck+06}) which are preferentially detected
in the radio and X-ray bands thanks to their luminous non-thermal
emission.  Since SNe with low mass loss rates generally give rise to
weak radio signals, they are only detectable nearby ($d\lesssim 30$
Mpc) with current cm-band facilities.  This, together with the discovery
rate of SNe Ibc within $d\approx 10$ Mpc (a few each decade),
statistically suggests that SN\,2007gr represents one of
the most ordinary SNe Ibc studied to date, characterized by explosion
and environmental properties that are typical of the bulk population.
With the significant improvement in continuum sensitivity enabled by
the Expanded VLA (EVLA; \citealt{pnj+09}), we will soon be able to
detect such ordinary SN Ibc to distances of $\sim 100$ Mpc which
will. in turn, broaden our understanding of these unique cosmic
explosions.

\medskip

The authors especially thank Mark Reid for helpful discussions.  We also thank
Dale Frail, Edo Berger, Andrew MacFadyen and Eli Waxman.  AMS is
supported by a Hubble fellowship. RAC acknowledges support from NASA
grant NNG06GJ33G.

\bibliographystyle{apj1b}

\clearpage

\begin{deluxetable}{lrrrrccc}
\tablecaption{VLA observations of SN\,2007gr}
\tablewidth{0pt}
\tablehead{
\colhead{Date} & \colhead{$F_{\nu,1.4}$} & \colhead{$F_{\nu,4.9}$} & \colhead{$F_{\nu,8.5}$} & \colhead{$F_{\nu,22}$} & \colhead{$\alpha_{2000}$ (8.5 GHz)} & \colhead{$\delta_{2000}$ (8.5 GHz)} & \colhead{Array}\\
\colhead{(UT)} & \colhead{($\mu$Jy)} & \colhead{($\mu$Jy)}
 & \colhead{($\mu$Jy)} & \colhead{($\mu$Jy)} & \colhead{02$^h$43$^m$+(s)} & \colhead{+37$^{o}$20'+(arsec)} & \colhead{Config.}
}
\startdata
2007 Aug 17.4 & \nodata & $777\pm 52$ & $659\pm 44$ & \nodata  & 27.9721$\pm$0.0007 & 44.699$\pm$0.006 & A\\
2007 Aug 18.4 & $123\pm 67$ & $954\pm 60$ & $640\pm 52$ & \nodata & 27.9696$\pm$0.0008 & 44.701$\pm$0.008 & A\\
2007 Aug 19.5 & \nodata & \nodata & \nodata & $< 327$ & \nodata & \nodata & A\\ 
2007 Aug 21.4 & $142\pm 47$ & $687\pm 58$ & $439\pm 48$ & \nodata & 27.9687$\pm$0.0011 & 44.674$\pm$0.009 & A\\
2007 Aug 24.5 & $< 228$ & $763\pm 64$ & $303\pm 56$ & \nodata & 27.9718$\pm$0.0017 & 44.653$\pm$0.018 & A\\
2007 Aug 30.6 & $677\pm 42$ & $417\pm 44$ & $172\pm 42$ & \nodata & 27.9751$\pm$0.0026 & 44.696$\pm$0.021 & A\\
2007 Sep 12.3 & $628\pm 86$ & $233\pm 76$ & $< 218$ & \nodata & \nodata & \nodata & A\\
2007 Sep 21.4 & $478\pm 159$ & $< 238$ & \nodata & \nodata & \nodata & \nodata & BnA\\
2007 Sep 30.6 & $529\pm 77$ & \nodata & \nodata & \nodata & \nodata & \nodata & BnA\\
2007 Oct 23.4 & $467\pm 60$ & $163\pm 45$ & \nodata & \nodata & \nodata & \nodata & B\\
2007 Nov 18.3 & $214\pm 42$ & $146\pm 34$ & \nodata & \nodata & \nodata & \nodata & B \\
2007 Dec 21.1 & $< 146$ & $< 157$ & \nodata & \nodata & \nodata & \nodata & B\\
2008 Jan 6.1 & $145\pm 27$ & \nodata & \nodata & \nodata & \nodata & \nodata & B\\
\enddata
\tablecomments{Upper limits are $3\sigma$.}
\label{tab:vla}
\end{deluxetable}

\clearpage

\clearpage

\clearpage

\clearpage

\clearpage

\end{document}